\documentclass[12pt]{article}
 \begin{document}
\pagenumbering{arabic}

 \medskip \begin{center} \Large {\bf Physics and Geometry\footnote{This paper is based on
 invited talks given by the author at the Gunnar Nordstr\"{o}m Symposium on Theoretical
 Physics at the University of
 Helsinki, Helsinki,
 Finland on August 28, 2003 and at the Freydoon Mansouri Memorial Session of
 the 3rd International Symposium on Quantum Theory and Symmetries at the University
 of Cincinnati, Cincinnati,
 OH, on September 13, 2003}}\\
  \normalsize                                 
 \bigskip \bigskip                   Peter G.O. Freund\footnote{freund@theory.uchicago.edu}\\
 \medskip \em{Enrico Fermi Institute and Department of Physics\\
 University of Chicago, Chicago, IL 60637}                                 
 							
 \bigskip \bigskip \bigskip
 
 \bf{Abstract}\\
 \end{center}
 
 Our understanding of the four basic concepts of Physics --- space, time, matter 
 and force --- has undergone radical change in the course of work on 
 unification, starting with 
 Maxwell's unification of electricity with magnetism, all the way to present day
 string theory. 
 What started
 as four independent concepts, with space and time postulated and the possible forms
 of matter and force arbitrarily chosen, now appear as different aspects of a rich and
 novel dynamically determined structure.

 \newpage

 Galileo and Newton, the founding fathers of modern Physics, have handed down to us
 four concepts: space, time, matter and force, in terms of which all of Physics
 is formulated. In terms of these concepts they set down what could now rightly be called
 the Galileo-Newton standard model of the seventeenth century. It involved
 
 --- a one-dimensional time continuum, 
 
 --- a three dimensional commutative Euclidean space, 
 
 --- arbitrarily chosen  matter, and
 
 --- arbitrarily chosen forces\\
all constrained through Newton's Galilei-invariant equations of motion.
This model was very close to and in agreement with experiment.

Over the past three decades, a lot has been achieved in unifying these four concepts, to
the point that now they appear as different aspects of one unifying fundamental concept.
Here I wish to review these recent developments and to do so I will first briefly
present some prerequisite older ideas.

Let me start with the concept of space. In its oldest form it was introduced axiomatically
by the ancient Greeks as a two- or three-dimensional Euclidean space. This form
involved the famous axiom of parallels, and was thought to be the only
possible space for over two millenia. It underwent a first major revision at the hands
of Bolyai and Lobachevsky in the nineteenth century. They discovered that
certain symmetric spaces support a geometry which satisfies all of Euclid's
axioms with the exception of the axiom of parallels. This revolutionary discovery
ultimately led to Felix Klein's Erlangen Program, in which geometry is intrinsically
related to group theory. According to Klein, if a group $G$  acts on a space $S$, then
the {\em geometry of} $G$ on $S$ is the study of $G$-invariant properties of the
"figures" of $S$. For Klein $G$ was to be a Lie group. For instance 3-dimensional Euclidean
geometry is recovered by choosing $G=E_3$, the three-dimensional Euclidean group
of translations and rotations. But this idea generalizes to any group, even to finite
groups.
We can speak of the geometry of a square or of a triangle, whose groups are obviously
finite. Equally well, we can associate a geometry to less trivial finite groups,
such as the 26 sporadic groups, even to the largest of these, the monster. This
is the outcome of the work of Bueckenhout \cite{FB}.  

By contrast, Riemann proposed another way of going beyond Euclidean geometry, in which
not the presence of a symmetry group, but that of a metric was to be the guiding
principle. The spaces discovered by Riemann, are not symmetric spaces in general.
At first sight they seem to defy the Erlangen program. The Riemannian and Kleinian
ideas of what a geometry should be were finally reconciled through the introduction
of the concept of a {\em connection} and the parallel transport it gives rise to.
The symmetry is then discovered to reside in the tangent spaces of the Riemannian space.

Moving on to Physics, one first adds time as a fourth dimension, to obtain a
4-dimensional Minkowski space in the absence of gravity, or a full-fledged
Riemann space upon the inclusion of gravity. The details of this Riemannian geometry
are then determined
--- {\em not} postulated --- through Einstein's equations from a knowledge of the otherwise
arbitrary distribution of matter. This arbitrariness is the same as the one
encountered already in the Galileo-Newton standard model, but here its appearance
is much more jarring. The point --- realized by Einstein already --- is that something 
geometric, the Einstein tensor $G_{\mu\nu}=R_{\mu\nu}-\frac{1}{2} g_{\mu\nu}$ has
to be proportional to something non-geometric, the energy-momentum
tensor $\Theta_{\mu\nu}$.

This difficulty can be cured by turning the energy-momentum tensor itself
into a geometric
object. But, along with bosonic scalar and vector fields, this tensor also involves
spinor fields, which obey Fermi-Dirac-Jordan
statistics and this leads to the introduction of transformations which mix Fermi and Bose
fields. This way we automatically land at the doorstep of {\em supersymmetry}. 

Remarkably, by the time we require of this supersymmetry 

i) that it contain
the 4-dimensional Poincar{\'e} algebra

ii) that it respect
the spin-statistics connection, and

iii) that it have nontrivial representations containing no fields with spin larger
than two,\\
there are
eight possible choices for it. They
are labeled by an integer $\cal{N}$, in terms of which the
number of Fermi generators of the supersymmetry algebra is $4\cal{N}$. In the maximal case 
$\cal{N} =  $ 8, there is a unique representation without spins larger than two, and all
basic fields must fit into it. This is a remarkable property, for this model fully
determines all possible forces and forms of matter, thus getting rid of the total
arbitrariness concerning their choices encountered in the Galileo-Newton standard model.
This is a major conceptual advance over the standard model of Particle Physics,
where the choice of the gauge group and of the representations to which the matter
fields must belong is also fraught with a great deal of arbitrariness, as
the only constraints on this choice come from anomaly cancellation conditions. 

The trouble with this $\cal{N} =  $ 8 supergravity in 4 dimensions is twofold. On
phenomenological level it does not agree with experiment. On the conceptual level,
its lagrangian, though well-determined, is very complicated and frustratingly
unilluminating. 

While the the phenomenological problem remains a serious obstacle, the conceptual problem
can be considerably alleviated. Maybe the theory looks so complicated because we are looking
at it in 4 dimensions, and this is in some way unnatural for it. Maybe its natural habitat
is in a higher dimensional space. This brings us to a Nordstr{\"o}m-Kaluza-Klein (NKK)type
approach to supergravity \cite{ACF},
and requires investigating the possible supergravities in higher
dimensions. In the supersymmetric context NKK theory is much more constrained than in the
non-supersymmetric case, in which the higher dimension can be arbitrary. Supergravities
only exist in dimensions $d \leq 11$. This is the counterpart in NKK theory of the
$\cal{N} \leq$ 8 of $d=4$ supergravities. It is essentially due to the requirement
that in a
supersymmetric theory the number $n_F$ of Fermi degrees of freedom must be equal to the
number $n_B$ of Bose
degrees of freedom. But $n_B$ increases polynomially with the dimension $d$ of
space-time, while $n_F$ increases exponentially with $d$, so that as $d$ increases,
$n_B$ cannot keep pace with $n_F$ and beyond a maximal dimension $d_{max}=11,$
there are no supergravities. So we find a maximal 11-dimensional supergravity, and it is
reasonable to investigate it.

The main surprise is that, while ordinarily we view the four-dimensionality of space-time
as a given, a choice from infinitely many possibilities, in the supersymmetric case there
are only eleven possible choices, and thereby we are presented with the opportunity
of predicting, or equivalently of understanding the criteria for choosing, the
dimension of space-time. 

This maximal supergravity $SUGRA_{11}$ is simple and compelling. It contains a graviton,
a gravitino and a a rank-three antisymmetric tensor field (3-form), all massless.
Under gauge transformations. the 3-form transforms as a potential. Its curl, a 4-form, is
a gauge invariant "field-strength." In the NKK spirit we now have
to see how this theory compactifies to lower dimensions.
But unlike NKK, we do not wish to postulate such a compactification, but rather
derive it dynamically. To this end, we have to find
classical solutions in which such compactification takes place. The simplest such
solution sets the gravitino field to zero and
identifies the field-strength 4-form with the volume form of a --- then
necessarily 4-dimensional --- submanifold $M_{4}$ of the
11-dimensional space-time manifold $M_{11}$. The Einstein equations then require
the structure of $M_{11}$ to be $M_{11} = M_4 \times M_7$. with $M_4$ and $M_7$,
both Einstein spaces, whose cosmological constants have opposite signs. Depending on which
of the two has positive cosmological constant, the 7- or the 4-dimensional
submanifold is the compact one. In the maximally symmetric case we thus
obtain a compactification to $AdS_4 \times S^7$, or $AdS_7 \times S^4$. It is interesting
that these compactifications prefer certain dimensions for the non-compact space, and
that one of these two preferred dimensions is four, which is "experimentally" viable.

There are other problems though. The small size of the compact manifold dictates a very
large (absolute value of the) cosmological constant of the non-compact manifold.
Moreover, the 4-dimensional particle spectrum is non-chiral, which again runs against the
experimental evidence. However recently it has been shown \cite{ADHL}
that this is no longer the case if the compact 7-manifold has singularities, e.g. conical
ones. In any case, these solutions will play an important role in what follows.

At a deeper level, this field theory is non-renormalizable, and as in the case of
Einstein gravity, this calls for drastic modifications. The problem is that the point-like
interaction vertices, allow for a precisely determined interaction event, which in turn
leads to bad ultraviolet behavior and results in non-renormalizability. By moving on
from a theory with interaction of point-like objects, to a theory in which the interacting
objects are extended, the interaction gets smeared out and the ultraviolet behavior
improves to the point that the theory becomes finite. We are thus led to strings
\cite{STRING}. Unlike
theories of interacting point particles, interacting strings automatically dial certain
"critical" dimensions, in which they can avoid the conformal anomaly. For bosonic strings
this critical dimension is 26, whereas for superstrings it is 10. The bosonic strings 
exhibit a tachyon instability. The tachyon is eliminated in the superstring
case. So, we end up in 10 dimensions, one dimension short of that of maximal supergravity.
One could take a cynical attitude to this fact, after all we are talking only of a ten
percent "correction" to the dimensionality of space-time. Yet in further developments
the theory will find its way back into 11 dimensions, as we shall see.

More than one superstring theory exists. There are the

--- open and closed type I superstrings,

--- closed type IIA superstrings,

--- closed type IIB superstrings,

--- closed heterotic $SO(32))$ superstrings, and

--- closed heterotic $E_8 \times E_8$ superstrings.\\

At first sight this seems disappointing, for if string theory is {\em the} ultimate
physical theory, it should rightfully be unique. But it was soon realized that these
five, on the face of it, different string theories are really but different aspects
of one overall theory, and as such are connected by what are known as dualities.

There are various types of dualities, but I will concentrate here on the particular
case of T-duality, and for simplicity I will consider the {\em closed} Bose string.
Then the critical dimension, as was already said, is $d_c = 26$. Of the 25 space dimensions
let us compactify one, say the 25-th, on a circle of radius $R$. The closed string
can then {\em wind}
on the compactification circle. The spectrum is
$$
m^2 = (\frac{n}{R})^2 + (\frac{wR}{\alpha'})^2 +\frac{2}{\alpha'}(N + \bar{N} -2). 
$$
The three terms here
correspond to the NKK modes, the winding modes and the usual closed string
oscillator modes respectively. This $m^2$ is invariant under the the replacements
$$
R \longleftrightarrow \frac{\alpha'}{R}  ~~~~~~~~~~~ n \longleftrightarrow w
$$
and so are the interactions. A string theory is then {\em dual} to another string theory
with a different
compactification radius, as required by this relation, and with the NKK and winding modes
interchanged. Notice that for the special radius $R_{SD} = \sqrt \alpha'$, the theory is
self-dual. This self-dual radius plays the role of a minimal length in the theory, for
as the compactification radius keeps decreasing below the value $R_{SD}$,
the compactification
radius of the dual (and therefore equivalent) theory keeps increasing above the
value $R_{SD}$. By the time the original compactification radius goes to zero, the dual
one goes to infinity. So we can shrink the size of the 25-th dimension up to $R_{SD}$, but
not beyond it, and we can not descend from 26 to 25 dimensions.
The closed string stays in 26
dimensions.

Things change for the open string, for which there is obviously no winding number and
therefore no exchange between NKK and winding modes can be envisioned. Therefore, there is
no minimal length and letting the compactification radius shrink to zero, the 26-dimensional
theory {\em does} reduce to a 25-dimensional one. But open and closed strings are made
of the same stuff. This is clear, for just as one end of one string can attach itself
to one end of another string to form a single open string at an open string vertex, so one
end of one string can attach itself to the other end of the {\em same} string, to form a
closed string. But as the compactification radius of one space dimension goes to zero,
closed strings are trapped in 26-dimensional space-time, and therefore so is the stuff
they are made of. It then stands to reason, that open strings experience the reduction
of the space dimension only through their ends, which therefore must be confined to a
25-dimensional subspace of 26-dimensional space-time. This 25-dimensional subspace is
called a D24-brane on account of the Dirichlet boundary conditions in $x^{25}$.

The D-branes are solitons and exist for superstrings as well. For type IIA
superstrings there exist D0-branes,
point-solitons on which strings can end. Their mass is 
$$
  m_{D0}=\frac{1}{g \sqrt \alpha'},
$$
where $g$ is the string coupling. What is more remarkable is the existence of n D0-brane
bound state configurations of mass $n \frac{1}{g \sqrt \alpha'}$. These states can in turn
be viewed as
spanning an NKK tower corresponding to a compactification of radius $g \sqrt \alpha'$.
As the coupling $g$ increases indefinitely, these states end up spanning a continuum. as if the
string had caused space to "grow" a new dimension. Superstrings started in a
10-dimensional space-time, so that when they grow this extra dimension, we land in an
11-dimensional {\em M-theory}. Its low energy limit is the maximal 11-d SUGRA which we
discussed earlier.

The D0-branes act as partons of this M-theory. Being zero-dimensional, they obey their
own quantum mechanics, a dimensionally reduced --- all the way to one dimension ---
form of 10-dimensional supersymmetric Yang-Mills theory.
The coordinates of this quantum mechanics are then matrices, which
in general do not commute. At small distances geometry then becomes {\em non-commutative}
\`{a} la Connes. 

I shall not discuss here the
phenomenologically interesting
string theory compactifications on 6-dimensional Calabi-Yau
manifolds. Rather I will consider strings on $AdS_n \times K_{10-n}$,
or M-theory on $AdS_n \times K_{11-n}$ background geometries, where
$K_m$ are compact $m$-dimensional manifolds. As discovered by Maldacena,
this leads to a remarkable holographic connection with a conformal
field theory (CFT) on
$(n-1)$-dimensional Minkowski space
$M_{n-1}$, the boundary of n-dimensional anti-de-Sitter space $AdS_n$. 

An idea of this AdS/CFT correspondence is most readily obtained, by returning to the 
hadronic strings from which string theory got its start. These hadronic strings
were abandoned in the wake of the great success of QCD. But here they re-emerge from
QCD, which plays the role of CFT in the AdS/CFT correspondence.

To simplify matters as much as possible, let me consider maximally supersymmetric
${\cal N}=4$ QCD, on 4-dimensional Minkowski space
$M_4$ with gauge group $G=U(N)$.
This theory is conformally symmetric: its couplings do not run, by default as
it were. The 4-dimensional conformal symmetry of this theory is $SO(4,2)$, which is
locally isomorphic to $SU(2,2)$. This theory also has a global $SO(6)$ symmetry,
which is in turn
locally isomorphic to $SU(4)$. The full superconformal symmetry is $SU(2,2|4)$.

For the string description, $SO(4,2)$ suggests an $AdS_5$ component
to the background geometry. It is, after all, the isometry group of $AdS_5$. Similarly,
$SO(6)$ is the isometry group of the 5-sphere, thus suggesting an overall
$AdS_5 \times S^5$ background geometry. This geometry is the vacuum of the
10-dimensional IIB SUGRA, which
is obtained in the same way as the $AdS_4 \times S^7$ vacuum of 11-dimensional SUGRA
dicussed above. That it is now a 5- and not a 4-dimensional AdS space, is due to the
simple fact that in IIB SUGRA, supersymmetry dictates the replacement
of the 4-form field strength encountered in 11-dimensional
SUGRA by a 5-form. So, the symmetries and supersymmetries of a string theory on an
$AdS_5 \times S^5$ background and of a maximally supersymmetric gauge theory on
4-dimensional Minkowski space coincide. There is more to this, and in fact the two theories
are equivalent: the Green functions of one can be obtained from those of the other.
If both $N$ --- which appears in the gauge group $SU(N)$ --- and its product with
the square
of the Yang-Mills coupling constant become very large, the correspondence becomes one
between the gauge theory and the IIB SUGRA.

In either case, this result is most surprising, for it states that the Physics of
a 4-dimensional quantum gauge field theory without gravity is the same as the Physics of
a 10-dimensional theory {\em with} gravity, be that theory a string theory or, in
the appropriate limit, a SUGRA.
This is the {\em AdS/CFT correspondence}. The very presence of gravity and
the dimension of space-time have become "relative": they depend on the description
we choose.  

As stated at the beginning of this paper, our picture of the four basic
concepts of Physics has undergone a major revision.

The old seventeenth century
Galileo-Newton standard model postulated a universal time, a 3-dimensional Euclidean
commutative space, and arbitrary forms of matter moving in it under the influence
of arbitrary forces. To its credit, this model was very close to experiment.

By contrast, at the geometric level
the twenty-first century string-theoretic unified theory presents us with a
$(1+3+\delta)$-dimensional non-Euclidean, and in general non-commutative space-time,
in which
the number of extra space dimensions is "predicted" to obey $\delta \leq 7$. The precise
dimension is "relative," it can change with the chosen description among
holgraphically dual pairs. All forces are now unified, as are all forms of matter.
Force and matter are themselves just different aspects of one and the same
agency, the string, and they also determine the geometry, which unlike Galileo-Newton, is
no longer postulated. In fact force, matter space and time, all the four basic concepts,
now determine each other and we face a unified whole. To {\em its} credit, this
theory of everything is very close to Mathematics.

 \end{document}